# Dissociation of NaCl in supercritical aqueous fluids of moderate and high concentrations: A molecular dynamics study

**Mikhail V. Ivanov[1,a)] and Olga V. Alexandrovich[1]**

**AFFILIATIONS**

[1]Institute of Precambrian Geology and Geochronology, Russian Academy of Sciences, Nab. Makarova 2, 199034, St. Petersburg, Russia

[a)]**Author to whom any correspondence should be addressed:** mikhail.v.ivanov52@gmail.com

**ORCID iDs**

Mikhail V. Ivanov 0000-0001-9780-3253

Olga V. Alexandrovich 0000-0003-2812-5393

**Abstract**

We report classical molecular dynamics simulations of NaCl association and dissociation in supercritical aqueous fluids over a wide range of salt concentrations, from moderate salinity to highly concentrated $H_2O$–NaCl mixtures attainable at high temperatures. The degree of dissociation α and the corresponding ideal dissociation constant $K_d$, derived directly from $\alpha$, were calculated as functions of the stoichiometric NaCl mole fraction at selected pressure–temperature (PT) conditions from 673.15 to 1273.15 K and from 0.1 to 2 GPa. At moderate salinity corresponding to a molality of approximately 1 mol/kg, NaCl remains largely dissociated ($\alpha \approx 0.3$–$0.7$ depending on pressure and temperature). In contrast, when the mole fraction of NaCl increases up to $x_{NaCl} = 0.333$ (27.8 mol/kg), the degree of dissociation tends towards zero, and most ions form $Na^+Cl^-$ contact pairs and multi-ion clusters. As a result of these competing trends, the mole fraction of structurally dissociated $Na^+$ and $Cl^-$ ions is a non-monotonic function of the stoichiometric NaCl concentration and typically reaches a maximum at $x_{NaCl} \approx 0.06$–$0.10$. This result shows that increasing salinity does not necessarily increase the abundance of structurally available chloride ions in supercritical aqueous fluids. Additional fixed-density simulations at 1 and 7 mol/kg extend the analysis up to 1673.15 K and separate the effects of temperature and density on the associate/dissociate state of the ions. The obtained concentration dependences provide molecular-level constraints for thermodynamic descriptions of concentrated supercritical electrolytes and for evaluating chloride availability in high-temperature aqueous fluids.

## 1. Introduction

Hydrothermal saline fluids play a fundamental role in heat and mass transfer in the Earth's crust and upper mantle. Dissolution, transportation, and precipitation of substances by aqueous fluids are among the principal processes responsible for formation of the vast majority of ore deposits, including the most economically significant deposits of gold on Earth (Heinrich, 2007; Fu, Touret, 2014). The temperature and pressure in the Earth's mantle and most of the Earth's crust correspond to the supercritical state of water. A common feature of such fluids is their high salinity (Hedenquist and Lowenstern, 1994; Manning, 2018), which is very far from the salinity of the most frequently studied dilute solutions with a molality well below 1m. In some cases, the mole fraction of salt in the fluid can reach 0.4 (i.e. molality 37m) (Manning, 2018). The properties of the supercritical saline fluids differ significantly from those at lower temperatures. These differences influence most geological processes in the depths of the Earth (Manning, 2018). The transport capacity of hydrothermal fluids depends largely on the concentration of free chloride ions, since $Cl^-$ acts as a ligand capable of forming stable metal chloride complexes (Hedenquist and Lowenstern, 1994; Yardley and Bodnar, 2014; Fowler and Sherman, 2020; Polidori *et al.*, 2021). The main source of $Cl^-$ ions for geological aqueous fluids is dissolved chlorides, including NaCl. The latter is a common and, in many cases predominant, component of these fluids.

Under normal conditions, NaCl behaves as a strong electrolyte and is usually considered to be completely dissociated in aqueous solution. However, at elevated temperatures and pressures, especially under supercritical conditions, the physicochemical properties of water change significantly. Its density and dielectric constant decrease, and the network of hydrogen bonds gradually breaks down. As a result, the ability of water to stabilize separated ions is weakened, which promotes the formation of ion pairs and clusters (Oelkers and Helgeson, 1990; 1991; 1993; Sherman and Collings, 2002; Bondarenko et al., 2006; Fowler and Sherman, 2020) even in electrolytes that are generally considered strong. Under such conditions, the simplified picture of complete dissociation becomes inadequate, and the degree of ion association must be explicitly taken into account, since even moderate changes in the equilibrium between free ions, contact ion pairs, and larger ion aggregates can strongly influence both macroscopic transport properties and thermodynamic behavior of the fluid. Quantitative knowledge of dissociation equilibria in concentrated salt-bearing fluids is necessary for realistic modeling of deep fluid transport, fluid-rock interactions, and metal mobility in the Earth's crust and upper mantle. More generally, the partitioning of dissolved salt between free ions and bound species affects not only chemical reactivity, but also conductivity, local structure, and short-range order, which are important for thermodynamic and geochemical descriptions of high-temperature fluids.

Experimental studies conducted at high temperatures and pressures have long indicated measurable deviations from complete dissociation in aqueous NaCl solutions (Quist and Marshall, 1968; Ho et al., 1994; Bondarenko et al., 2006). More recent spectroscopic and conductivity measurements also indicate significant ion association under hydrothermal and supercritical conditions (Elbers et al., 2021; Sinmyo and Keppler, 2017; Guo and Keppler, 2019).

The properties of the supercritical aqueous fluids have been studied using molecular dynamics (MD) methods in many papers over several decades (Belonoshko and Saxena, 1991; Chialvo et al., 1995; Brodholt, 1998; Driesner et al., 1998; Koneshan and Rasaiah, 2000; Sherman and Collings, 2002; Chialvo and Simonson, 2003; Bondarenko et al., 2006; Lümmen and Kvamme, 2007; Yui et al., 2010; Timko et al., 2010; Orozco et al., 2014; Sakuma and Ichiki, 2016; Mei et al., 2018; Zhang et al., 2020; Fowler and Sherman, 2020; Elbers et al., 2021; Sahle et al., 2022; Song et al., 2022; Simeski and Ihme, 2023; Fowler et al., 2024; Duan et al., 2025). In the case of the NaCl–$H_2O$ system, molecular dynamics studies have focused mainly on the microscopic structure, including hydration shells, pair distribution functions, and short-range correlations between ions. Considerable attention was also paid to the density of the system. Several studies have provided qualitative evidence for ion association and ion pair formation NaCl-containing fluids (Sherman and Collings, 2002; Zhang et al., 2020; Fowler and Sherman, 2020). In this regard, molecular dynamics simulations are particularly valuable because they allow one to distinguish between different associated states and relate macroscopic observables to microscopic configurations. Numerical values of the degree of dissociation $\alpha$ and the dissociation constant were obtained from molecular dynamics calculations (Mei et al., 2018) for HCl in an aqueous fluid with molality 1m. The association degree values for the NaCl–$H_2O$ system at different temperatures from 27°C to 600°C at a fixed molality (2.78 mol/kg under supercritical conditions) were obtained in work (Elbers et al., 2021). In this work, the association of $Na^+$ and $Cl^-$ ions was also experimentally studied. A high degree of association of $Na^+$ and $Cl^-$ ions in supercritical fluids was also found in the work (Bondarenko et al., 2006). In an MD study of the relatively dilute supercritical fluid NaCl–$H_2O$ (Zhang et al., 2020), preferred thermodynamic conditions for association of $Na^+$ and $Cl^-$ ions were found. The dependence of the properties of a supercritical NaCl aqueous fluid under high pressure on its molality in the range from 0.5m to 16m was studied in MD calculations (Sherman and Collings, 2002; Fowler and Sherman, 2020). These studies do not provide values for dissociation or association constants, but conclude that there is a strong association between $Na^+$ and $Cl^-$ ions.

The thermodynamic properties of supercritical aqueous saline fluids have been the subject of a number of theoretical works (Helgeson et al., 1981; Shock et al., 1992; Oscarson et al., 2001; 2004; Driesner, 2007; Sverjensky et al., 2014; Mao et al., 2015), including the HKF theory and more recent DEW model. The properties of supercritical aqueous fluids, including the degree of dissociation of dissolved electrolytes, depend largely on temperature and pressure, as well as on the composition of the fluid. Existing thermodynamic models of multicomponent supercritical aqueous fluids have commonly treated the degree of electrolyte dissociation using approximations that depend mainly on temperature and pressure, while its possible dependence on electrolyte concentration has usually remained outside the scope of these models (Aranovich et al., 2010; Ivanov and Bushmin, 2019, 2021; Ivanov, 2023; Makhluf et al., 2023). Obtaining concentration dependences of α is therefore important for further refinement of thermodynamic models of concentrated supercritical fluids.

The aim of our work is to study the dependence of the degree of dissociation of NaCl (or the degree of association of ion pairs $Na^{+}Cl^{-}$) on the concentration of salt, starting from moderate salinity of 1 mol/kg to high NaCl concentrations that can be achieved at high temperatures.

For our calculations, we chose classical molecular dynamics with potentials (Toukan and Rahman, 1985; Smith and Dang, 1994), which have been shown to give results close to experimental data. This approach requires fewer computational resources than computational approaches classified as *ab initio* MD. Although *ab initio* MD provides a more direct description of electronic structure, its computational cost limits systematic studies over broad ranges of temperature, pressure, and composition. Classical MD therefore remains a practical approach for obtaining statistically meaningful trends, provided that the interaction potentials are carefully chosen and validated. Our classical MD approach allowed us to obtain detailed data on the dissociation of NaCl over a wide range of temperatures, pressures, and NaCl concentrations in the supercritical fluid $H_2O$–NaCl.

## 2. Computational method

All our simulations were performed using the MDynaMix package (Lyubartsev and Laaksonen, 2000). Water molecules were described by the flexible SPC/fw model (Toukan and Rahman, 1985), while ion–ion and ion–water interactions were taken from the Smith–Dang force field (Smith and Dang, 1994). These models have been extensively applied to concentrated electrolyte systems under elevated PT conditions (Lyubartsev and Laaksonen, 1996; Sherman and Collings, 2002; Mirzoev and Lyubartsev, 2011) and have been shown to provide results close to experimental data.

The temperature was controlled using the Nosé thermostat. The simulation was carried out in cubic cells with periodic boundary conditions and sides of length $L_{box}$. In fixed pressure calculations, $L_{box}$ values were initially iteratively adjusted to match target pressure values. The long-range electrostatic interactions were treated by the Ewald method with the cutoff radius in the range from 0.33 $L_{box}$ to 0.5 $L_{box}$. The integration time step was set to 0.4 fs and was reduced as needed to ensure stability. After a 200 ps equilibration period, the simulation was run for periods ranging from 4 to 40 ns in order to ensure the required statistical accuracy.

The calculations for the molality close to $b = 1$ mol/kg (1m) were performed for a system containing 8 pairs $Na^+Cl^-$ (or stoichiometric 8 NaCl molecules) and 440 $H_2O$ molecules ($N_{NaCl} = 8$, $N_{H2O} = 440$). This size of the system is sufficient to obtain the decimal logarithm of the ideal dissociation constant $\log_{10}K_d$ with errors of the order 0.01 – 0.02. This fact was confirmed by test calculations for larger systems. For example, our calculations for the above system at $T = 1273.15$ K and $P = 1$ GPa give $\log_{10}K_d = -2.2535$, $\sigma = 0.007$. For a similar system of double size, i.e. with the stoichiometric composition of 16 molecules NaCl and 880 molecules $H_2O$, our result is $\log_{10}K_d = -2.2553$, $\sigma = 0.007$. This 8:440 ($x_{NaCl}$:$x_{H2O}$ = 1:55) system corresponds to the mole fraction of NaCl $x_{NaCl} = 0.018$. Here and in the following $x_{NaCl}$ denotes the stoichiometric mole fraction of the NaCl in the fluid; $x_{H2O}$ denotes the mole fraction of $H_2O$.

For larger mole fractions of NaCl, calculations were performed for the following systems:

- 16 NaCl : 512 $H_2O$ ($x_{NaCl} = 0.030$; $b = 1.73$ mol/kg),
- 16 NaCl : 368 $H_2O$ ($x_{NaCl} = 0.042$, $b = 2.41$ mol/kg),
- 32 NaCl : 512 $H_2O$ ($x_{NaCl} = 0.059$, $b = 3.47$ mol/kg),
- 32 NaCl : 352 $H_2O$ ($x_{NaCl} = 0.083$, $b = 5.05$ mol/kg),
- 32 NaCl : 256 $H_2O$ ($x_{NaCl} = 0.111$, $b = 6.94$ mol/kg, referred below to 7m),
- 64 NaCl : 384 $H_2O$ ($x_{NaCl} = 0.143$, $b = 9.25$ mol/kg),
- 64 NaCl : 256 $H_2O$ ($x_{NaCl} = 0.200$, $b = 13.9$ mol/kg),
- 128 NaCl : 384 $H_2O$ ($x_{NaCl} = 0.250$, $b = 18.5$ mol/kg),
- 128 NaCl : 256 $H_2O$ ($x_{NaCl} = 0.333$; $b = 27.8$ mol/kg).

Examples of the radial distribution functions $g(r_{Na^+Cl^-})$ for the distance $r_{Na^+Cl^-}$ between $Na^+$ and $Cl^-$ ions, obtained in our calculations are shown in the left panels of Fig. 1. All the radial distribution functions have a pronounced maximum near $r_{Na^+Cl^-} = 2.75$ Å. This distance 2.75 Å is slightly less than the distance 2.82 Å between Na and Cl atoms in the NaCl crystals. The functions shown in the left panels of Fig. 1 reflect the distribution of all the ions with a charge opposite to the central ion. Therefore, the distributions exhibit tails corresponding to all the ions of opposite charge located further away than the one closest to the central ion. The relative amplitude of these tails compared to the first peak increases with increasing the salt concentration.

To obtain information about the association/dissociation state of the central ion, radial distribution functions that take into account only the nearest ion of the opposite charge $g_1(r_{Na^+Cl^-})$ are more informative. These functions are shown in the right panels of Fig. 1. Increasing the mole fraction of NaCl reduces the probability that an ion with the opposite charge will not be detected near the central ion. The second maximum in the distribution is visible in Fig. 1(b). A very short tail is present in Fig. 1(d). The spatial density of ions does not provide sufficient space free of ions of opposite charge at the ratio $x_{NaCl}$:$x_{H2O}$ = 1:2, as seen in Fig. 1(f).

It is natural to relate the associated state of $Na^+Cl^-$ to the first maximum of the radial distribution function in vicinity of $r_{Na^+Cl^-} = 2.75$ Å, and to establish the boundary between the associated and dissociated states at the first minimum of this function. At a certain critical distance q, we consider the $Na^+Cl^-$ pair with $r_{Na^+Cl^-} > q$ to be dissociated, and in other cases, associated. In our calculations of the degree of dissociation $\alpha$, we used $q = 3.72$ Å. This value is equal to a distance of closed approach of ions $Na^+$ and $Cl^-$ proposed by Helgeson et al. (1981) for the HKF-type activity-coefficient calculations. Small deviations in the value of $q$ result in minor changes in the numerical results presented below. The areas under the radial distribution function graphs, corresponding to the associated $Na^+Cl^-$, are shaded in Fig. 1.

The distinction between associated and dissociated states, based on the distance of the $Cl^-$ ion from the nearest $Na^+$ ion, is consistent with the role of $Cl^-$ ions in the mobility of metals in the Earth's crust and mantle. A $Cl^-$ ion sufficiently far from $Na^+$ ions can form metal-chloride complexes. In contrast, the electrostatic field of the $Cl^-$ ion with a neighboring $Na^+$ ion decreases much faster with increasing distance than the field of an isolated chloride ion. This can hinder the formation of stable metal-chloride complexes. Since our focus is on the structural availability of dissociated chloride ions for complex formation, we use a geometric criterion rather than a residence time-based definition of the ion pairing (Marcus and Hefter, 2006). In terms of widely used classifications of ion pairs and solvent-separated configurations, our definition of the associated $Na^+Cl^-$ state is closest to that of a contact ion pair (CIP) (Marcus and Hefter, 2006; Zhang et al., 2020).

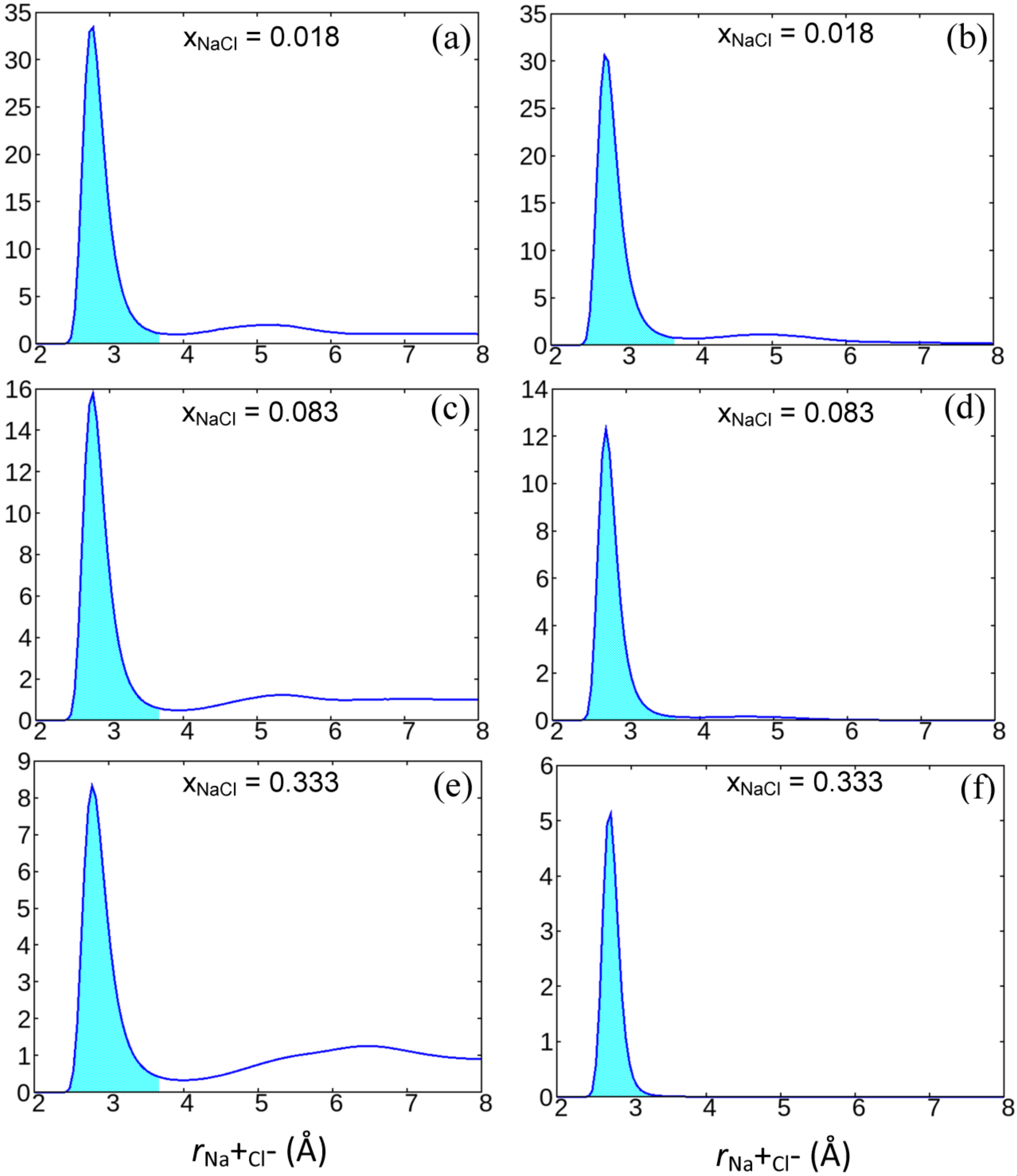

**FIG. 1.** Panels (a), (c), and (e) show the radial distribution functions of the distance between $Na^+$ and $Cl^-$ ions at $T$ = 673.15 K, $P$ = 0.1 GPa. Panels (b), (d), and (f) show the same functions for the distance to the nearest ion of opposite charge. The area corresponding to the bound state of the ions is shaded.

**Calculation of $\alpha$ and $K_d$.** During the numerical solution of the MD equations, the positions of the $Na^+$ and $Cl^-$ ions (ion configurations) were stored at regular intervals throughout the simulation, typically every Δts = 100 fs. The configurations obtained after the initial equilibration period were used to obtain values of the pressure (the pressure was calculated in MDynaMix from the atomic virial expression), the degree of dissociation, and the decimal logarithm of the ideal dissociation constant.

The average values of the degree of dissociation were obtained as follows. For each stored configuration $j$, the number of dissociated NaCl molecules $N_j^{\mathrm{diss}}$ was considered to be the number of ions of one charge (e.g., $Na^+$) that have no ions of opposite charge closer than $r_{\mathrm{Na^+Cl^-}} = q$. When the total number of stored configurations is $S_{\mathrm{tot}}$, the average degree of dissociation is

$$\alpha = \left( \sum_{j=1}^{S_{\mathrm{tot}}} N_j^{\mathrm{diss}} \right) / (N_{\mathrm{NaCl}} S_{\mathrm{tot}}) \qquad (1)$$

The distances between $Na^+$ and $Cl^-$ ions in successive configurations are not statistically independent. This complicates the estimation of the statistical error of $\alpha$. However, if the series of configurations is long enough to avoid statistical dependence of configurations separated by large periods of simulation time, one can divide the entire series of configurations into $n$ large parts and calculate the average degree of dissociation $\alpha_i, i = 1,\ldots,n$ independently for each part. For $n$ equal parts of the series, the error of $\alpha$, obtained on the whole series, can be estimated as

$$\sigma_\alpha = \left( \frac{1}{n-1} \sum_{i=1}^{n} (\alpha_i - \alpha)^2 \right)^{1/2} / \sqrt{n} \qquad (2)$$

The quantity in parentheses is the variance of $\alpha_i$, which is $n$ times greater than the variance of the entire series. The variations of $\sigma_\alpha$ with $n$ are insignificant for moderate values of $n$. All the values of $\sigma_\alpha$ considered in the calculations presented below were obtained with $n = 9$.

In addition to the average degree of dissociation, we also present the values of the decimal logarithm of the NaCl ideal dissociation constant $K_\mathrm{d}$ in the form (Wright, 2007)

$$K_\mathrm{d} = \frac{[\mathrm{Na}^+][\mathrm{Cl}^-]}{[\mathrm{NaCl}]_{\mathrm{ass}}} = \frac{\alpha^2}{1-\alpha} x_{\mathrm{NaCl}}, \qquad (3)$$

All the activity coefficients in equation (3) are assumed to be equal to unity. Due to the strong non-ideality of the system, $K_\mathrm{d}$ depends on $x_{\mathrm{NaCl}}$ and should not be confused with a thermodynamic standard-state dissociation constant. The logarithm of the quantity (3) is convenient for the numerical description and approximation data obtained for $\alpha$. The knowledge of $K_\mathrm{d}$ allows easily obtain the dissociation degree at given $P$, $T$, and $x_{\mathrm{NaCl}}$.

The average values of $\log_{10}K_\mathrm{d}$ were calculated from the average $\alpha$ via eq. (3). The errors $\sigma_{\log K}$ of $\log_{10}K_\mathrm{d}$ were calculated analogously to $\sigma_\alpha$. For all the results presented below, the MD simulation time was long enough to achieve the estimated errors $\sigma_{\log K} \leq 0.01$. The typical simulation times ranged from 4 to 40 ns. Thus, since $\Delta t_\mathrm{s} = 100$ fs, each obtained value of $\alpha$ or $\log_{10}K_\mathrm{d}$ is the result of averaging at least of 40,000 ion configuration samples. Strictly speaking, the samples are not completely independent, since they belong to the same long trajectory. However, the time interval between the samples $\Delta t_s$ was chosen to be large enough to exclude unnecessary samples that do not contain statistically significant information. We ran test calculations with a long total simulation times and recording samples every 4 fs. It was then possible to calculate $\alpha$ and $\log_{10}K_d$ and their statistical errors using the different $\Delta t_s$ between the samples included in the statistics. Variations in $\Delta t_s$ in the range from $\Delta t_s = 4$ fs to $\Delta t_s = 100$ fs had virtually no effect on the mean values and their statistical errors, despite the change in the number of the samples included in the statistics. On the other hand, choosing $\Delta t_s$ sufficiently greater than $\Delta t_s = 100$ fs led to an increase in statistical errors with increasing $\Delta t_s$ due to the corresponding decrease in the number of samples.

## 3. Results

### *3.1. Calculations at fixed pressures*

We performed calculations of the degree of dissociation of NaCl $\alpha$ and the decimal logarithm of the ideal dissociation constant $\log_{10}K_d$ at four temperatures: $T = 673.15$ K, $T = 873.15$ K, $T = 1073.15$ K, and $T = 1273.15$ K. For $T = 673.15$ K, the values of $\alpha$ and $\log_{10}K_d$ were obtained at pressures $P = 0.1$ GPa, $P = 0.2$ GPa, $P = 0.4$ GPa, and $P = 1$ GPa. For $T = 873.15$ K and $T = 1073.15$ K, the calculations were carried out at $P = 0.2$ GPa, $P = 0.4$ GPa, and $P = 1$ GPa. For $T = 1273.15$ K, the pressures $P = 0.2$ GPa, $P = 0.4$ GPa, $P = 1$ GPa, and $P = 2$ GPa were studied. The dependences of $\log_{10}K_d$ and $\alpha$ on the stoichiometric mole fraction of NaCl are presented in Fig. 2. At all the temperatures and pressures (except for a more complicated dependence at $T = 1273.15$ K and $P = 0.2$ GPa), a rapid decrease in the degree of dissociation is observed with an increase in the concentration of NaCl in the liquid.

Our $\alpha(x_{NaCl})$ dependences start at the molality of $b \approx 1$ mol/kg. At this stage, the values of $\alpha$ for most temperatures and pressures are about $\alpha = 0.5$–$0.7$ or at least $\alpha \approx 0.3$. At the maximal concentration of NaCl $x_{NaCl} = 0.333$ ($b = 27.8$ mol/kg), shown in Fig. 2, the degree of dissociation of NaCl is significantly less than 0.01. In terms of the ideal dissociation constant, this means a decrease in typical $\log_{10}K_d$ values from $\log_{10}K_d \approx -1.5, -2$ to values close to $\log_{10}K_d = -5$. All the results of the molecular dynamics simulation for $\log_{10}K_d$ as a function of $x_{NaCl}$ can be accurately approximated by the equation

$$\log_{10} K_d = a + b x_{NaCl} + c\sqrt{x_{NaCl}} \tag{4}$$

with three fitting parameters $a$, $b$, and $c$. These approximations are shown in Figs. 2(a), 2(c), 2(e), and 2(g) with curves. The fitting curves for $\alpha(x_{NaCl})$ in Figs. 2(b), 2(d), 2(f), and 2(h) are obtained from approximation (4) using the equation

$$\alpha = \frac{K_d}{2x_{NaCl}}\left(\sqrt{1 + 4x_{NaCl}/K_d} - 1\right), \tag{5}$$

which is a solution of equation (3) with respect to $\alpha$.

The sharp decrease in $\alpha$ with increasing $x_{NaCl}$ can be explained by simple geometric reasons. Knowing the volume of the cell, the number of the ions in it and assuming no interaction between the ions, it is easy to calculate the average distance between the ions $d_{ii}$. At $T = 1273.15$ K, $P = 2$ GPa, and $b = 1$ mol/kg, $d_{ii} = 9.43$ Å. At the same temperature and pressure, and at $b = 27.8$ mol/kg, $d_{ii} = 3.87$ Å. This means that in the statistical distribution, many ions will have their nearest neighbor located at a distance of less than 3.72 Å. Besides these geometric reasons, there are also attractive forces between ions of opposite charge. As a result of these two factors, the proportion of bound ions exceeds 99%.

Another example could be the fluid $H_2O$–NaCl at temperature $T = 1273.15$ K and pressure $P = 1$ GPa. Under these conditions, pure NaCl is in a molten state (Pistorius, 1966) and can be mixed with water in arbitrary proportions. At the last point $x_{NaCl} = 0.333$ shown in the graphs, the average distance between ions is $d_{ii} = 4.04$ Å. At $x_{NaCl} = 0.5$, that is, in a mixture of equal numbers of the $H_2O$ molecules and $Na^+Cl^-$ pairs, $d_{ii} = 3.66$ Å. This means that if the distribution of the ions were spatially uniform, there would be no dissociated ions. However, since a significant fraction of the ions form pairs in close-contact or multi-ion complexes, some ions remain dissociated, $\alpha = 0.009$, $\log_{10}K_d = -6.39$. At $x_{NaCl} = 0.667$, this is in fact a solution of $H_2O$ in NaCl, $d_{ii} = 3.43$ Å. This is a macroscopically homogeneous solution, but the same formation of contact pairs and multi-ion complexes in the molecular structure of the solution ensures a minimal proportion of dissociated $Na^+$ and $Cl^-$ ions. The degree of dissociation of NaCl is $\alpha = 0.001$, $\log_{10}K_d = -8.17$.

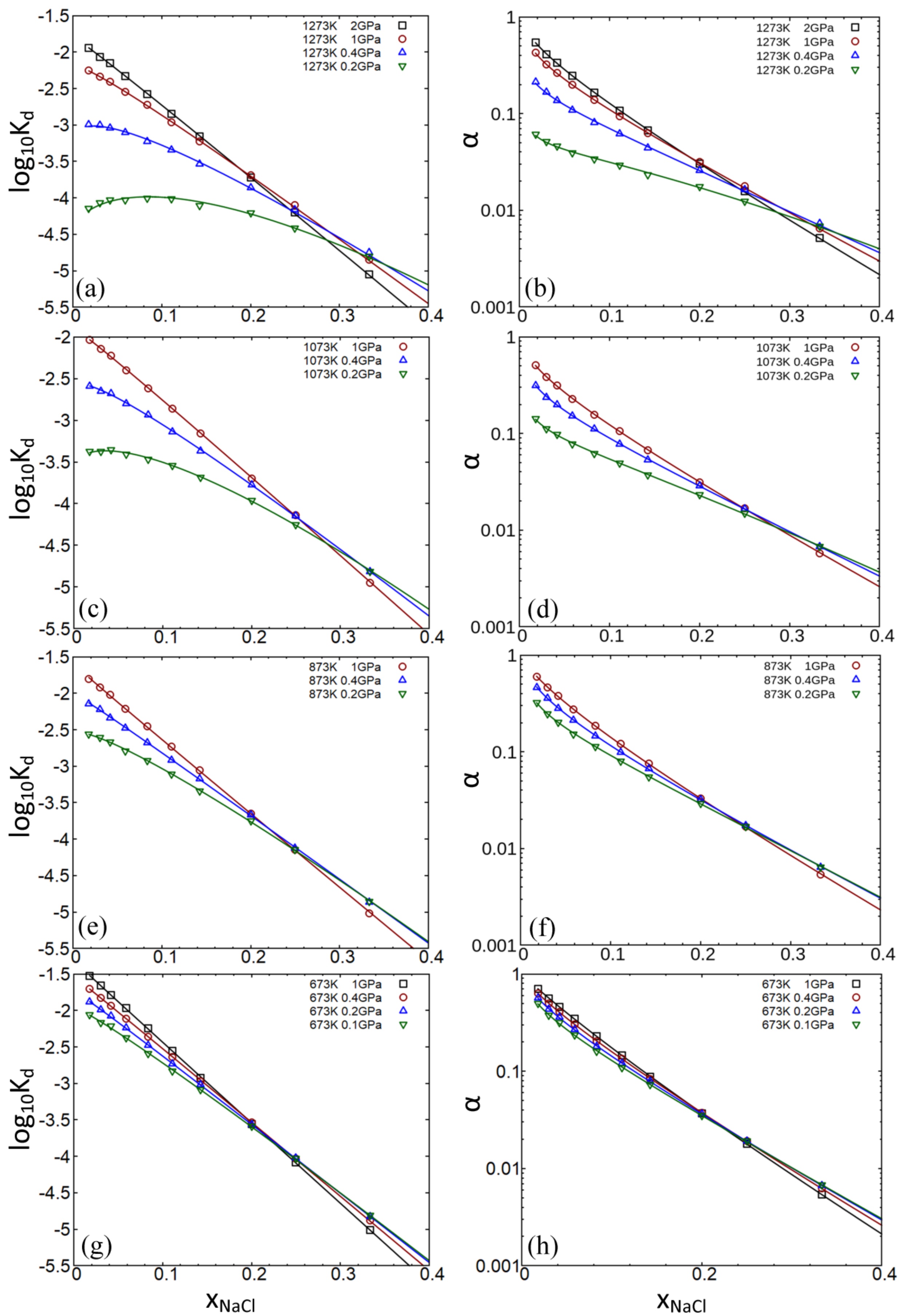

**FIG. 2.** Logarithm of the ideal dissociation constant and the degree of dissociation of NaCl as functions of the mole fraction of NaCl in the fluid at different temperatures and pressures.

At a constant temperature and moderate or not very high mole fractions of NaCl (up to $x_{NaCl} = 0.2$), the degree of dissociation and $\log_{10}K_d$ increase with increasing pressure. This effect is easily explained by the stronger interaction of water molecules with $Na^+$ and $Cl^-$ ions in denser fluids. On a macroscopic scale, this effect can be associated with an increase in the dielectric constant of water with an increase in its density. In the concentration range between $x_{NaCl} = 0.2$ and $x_{NaCl} = 0.25$, this dependence of $\alpha$ and $\log_{10}K_d$ on the pressure changes its sign. At the maximal value of $x_{NaCl} = 0.333$, the main tendency is to decrease $\alpha$ and $\log_{10}K_d$ with increasing $P$. In this range of NaCl concentrations, the geometric effect described above may be more significant.

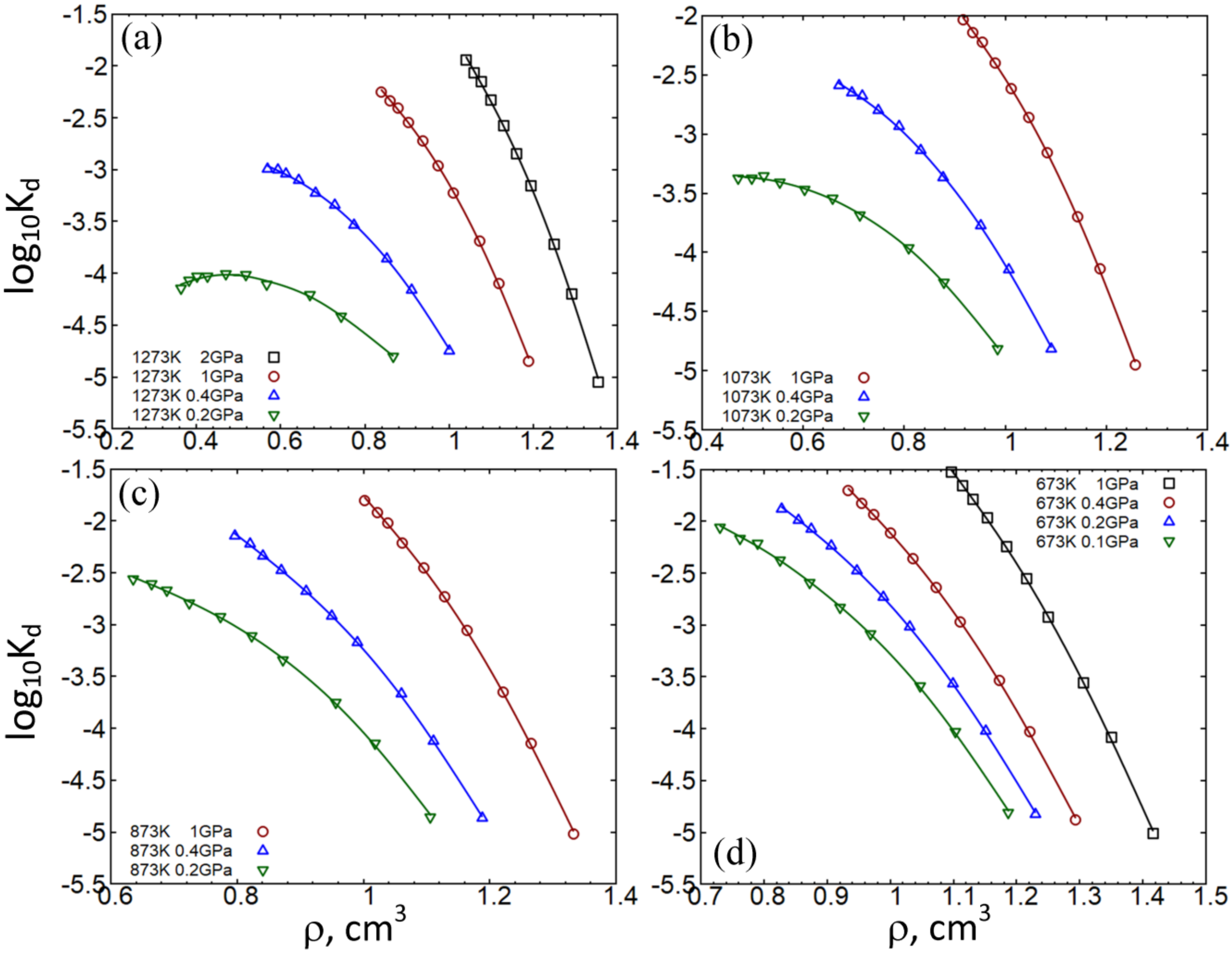

**FIG. 3.** Logarithm of the ideal dissociation constant of NaCl as a function of the density of the fluid.

The influence of the density of the fluid on the ideal dissociation constant is shown more clearly in Fig. 3. We present here the curves for $\log_{10}K_d$ as functions of the density $\rho$ of the fluid at fixed temperatures and pressures. In general, these curves are similar to corresponding curves in Fig. 2. The difference in their shape and relative position is due to the fact that the density is highly dependent on the pressure. As a result, the curves corresponding to different pressures are located far apart from each other, unlike the curves close together in Fig. 2. The effect of density at constant temperatures, discussed above, is clearly visible in Fig. 3. The position of the end points of the curves in Fig. 3 clearly illustrates the influence of density on the value of $\log_{10}K_d$ and degree of dissociation of NaCl. The upper ends of the curves correspond to the minimal mole fraction of NaCl $x_{NaCl} = 0.018$. At all the temperatures shown in Fig. 3, these points show an increase in $\log_{10}K_d$ with increasing pressure. The lower ends of the curves correspond to $x_{NaCl} = 0.333$. The main trend observed here is a decrease in $\log_{10}K_d$ with increasing pressure.

For three pressures, namely $P = 0.2$ GPa, $P = 0.4$ GPa, and $P = 1$ GPa, we performed calculations for all the temperatures studied. The temperature dependences of $\log_{10}K_d$ and $\alpha$ for several fixed NaCl concentrations are shown in Fig. 4. The temperature in this graph is shown in degrees Celsius. A common feature of these dependences is a decrease in the degree of dissociation and the ideal dissociation constant with increasing temperature. This trend is observed for all the curves at molality $b = 1$ mol/kg, $b = 3.5$ mol/kg, and $b = 7$ mol/kg. The only exception is the curve for $b = 14$ mol/kg and $P = 1$ GPa, which corresponds to a high proportion of associated ions and a high fluid density. The effect of decreasing the degree of dissociation with increasing temperature is partly caused by the decrease in the fluid density with increasing temperature at constant pressure. This effect should be especially pronounced at low pressure, which is observed in the curves for $P = 0.2$ GPa. On the other hand, this effect is also observed at constant densities. The corresponding results are presented in the following subsection.

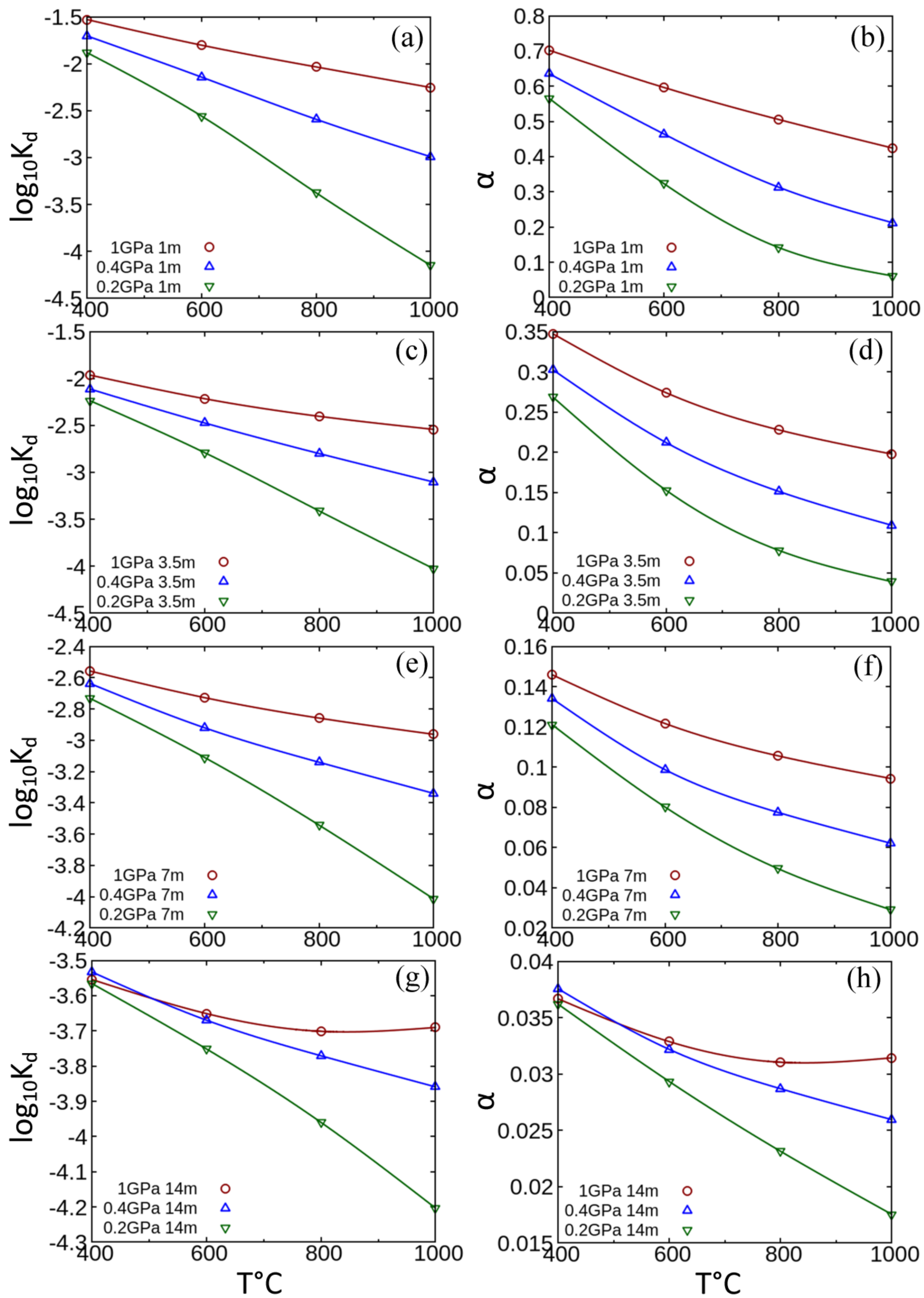

**FIG. 4.** Logarithm of the ideal dissociation constant and degree of dissociation of NaCl as a function of the temperature for a set of pressures and molality values.

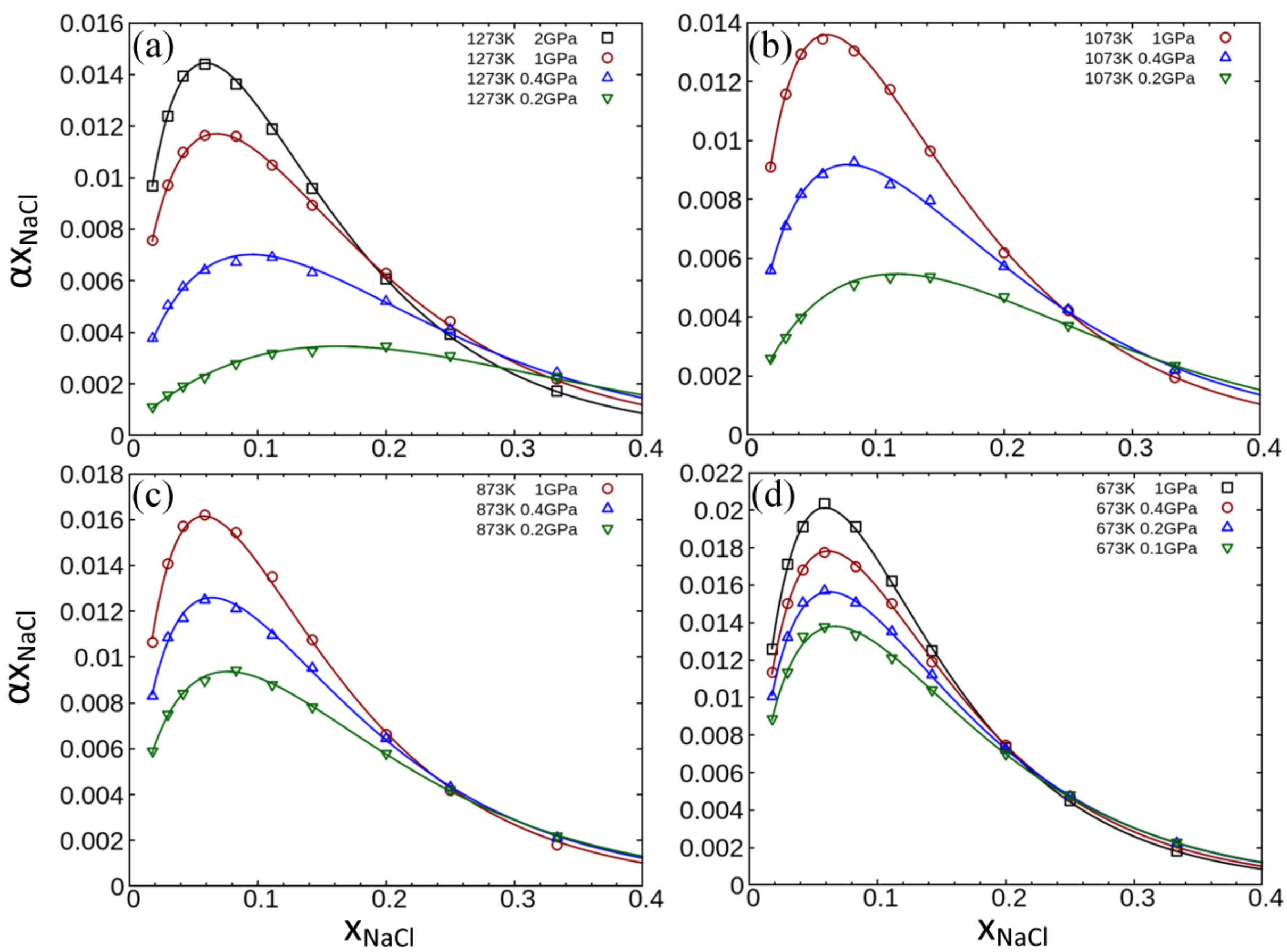

**FIG. 5.** Mole fraction of dissociated ion pairs as a function of the stoichiometric mole fraction of NaCl in the fluid.

The ability of fluids to dissolve metals and transport metal-chloride complexes formed in the Earth's crust and upper mantle depends critically on the concentration of free $Cl^-$ ions in the fluid (Fowler and Sherman, 2020; Manning, 2018). One stoichiometric mole of $H_2O$–NaCl fluid contains $\alpha x_{NaCl}$ moles of dissociated $Cl^-$ ions. It is appropriate to consider this value as the mole fraction of dissociated $Cl^-$ ions, that is, the ratio of the number of these ions to the stoichiometric number of $H_2O$ and NaCl molecules in the system. We will denote this quantity by $x_{Cl-}$. This quantity coincides with the analogous mole fraction of $Na^+$ ions $x_{Na+}$ and with the mole fraction of dissociated $Na^+Cl^-$ pairs. The value of $\alpha x_{NaCl}$ as a function of $x_{NaCl}$ is shown in Fig. 5. The maximum of the $x_{Cl-}(x_{NaCl})$ function indicates the concentration range in which the abundance of structurally available chloride ions is highest and, therefore, the range of the maximum capacity of the fluid to form and transport metal-chloride complexes. At all the temperatures studied, the curves corresponding to the highest pressure (and hence to the highest fluid density) reach their maximum value near $x_{NaCl} = 0.06$. At the lowest temperature $T = 673.15$ K, the curve for $P = 1$ GPa has its maximum almost exactly at $x_{NaCl} = 0.06$. The curves corresponding to lower pressure are located below the line corresponding to $P = 1$ GPa. The maxima of these dependences undergo a slight shift to the right as the pressure decreases. At $P = 0.1$ GPa, the function $x_{Cl-}(x_{NaCl})$ reaches its maximum value near $x_{NaCl} = 0.07$. At temperatures above $T = 673.15$ K, the picture is similar, but with increasing temperature, the separation between the curves for different pressures increases. This is mainly due to significant variations in fluid density depending on pressure. The effect of shifting the maximum of the function $x_{Cl-}(x_{NaCl})$ to the right also becomes more pronounced. At the lowest pressure $P = 0.2$ GPa in Figs. 3(a), 3(b), and 3(c), the maximum of $x_{Cl-}(x_{NaCl})$ is located at $x_{NaCl} \approx 0.08$ for $T = 873.15$ K, $x_{NaCl} \approx 0.10$ for $T = 1073.15$ K, and $x_{NaCl} \approx 0.16$ for $T = 1273.15$ K. The fluid density at $P = 0.2$ GPa and temperatures $T = 1073.15$ K and $T = 1273.15$ K are quite low. More representative curves could be those for $P = 0.4$ GPa with their maxima at $x_{NaCl} \approx 0.08$ ($T = 1073.15$ K) and $x_{NaCl} \approx 0.10$ ($T = 1273.15$ K). Thus, the highest abundance of dissociated chloride ions in supercritical $H_2O$–NaCl fluids is generally associated with NaCl mole fractions in the range $x_{NaCl} = 0.04 – 0.10$. This concentration range may therefore be especially favorable for metal mobilization, transport as metal-chloride complexes, and subsequent precipitation in ore-forming environments. Increasing the NaCl mole fraction above $x_{NaCl} = 0.10$ leads to a significant decrease in the fraction of dissociated chloride ions and, consequently, may reduce the chloride-related metal transport capacity of the fluid. At $x_{NaCl} = 0.333$, this reduction can reach approximately one order of magnitude.

In Fig. 6, the same quantity $\alpha x_{NaCl}$ is presented as a function of the fluid density. As in Fig. 3, the curves corresponding to different pressures are shifted horizontally due to the strong dependence of the density on the pressure. This effect is more pronounced at higher temperatures.

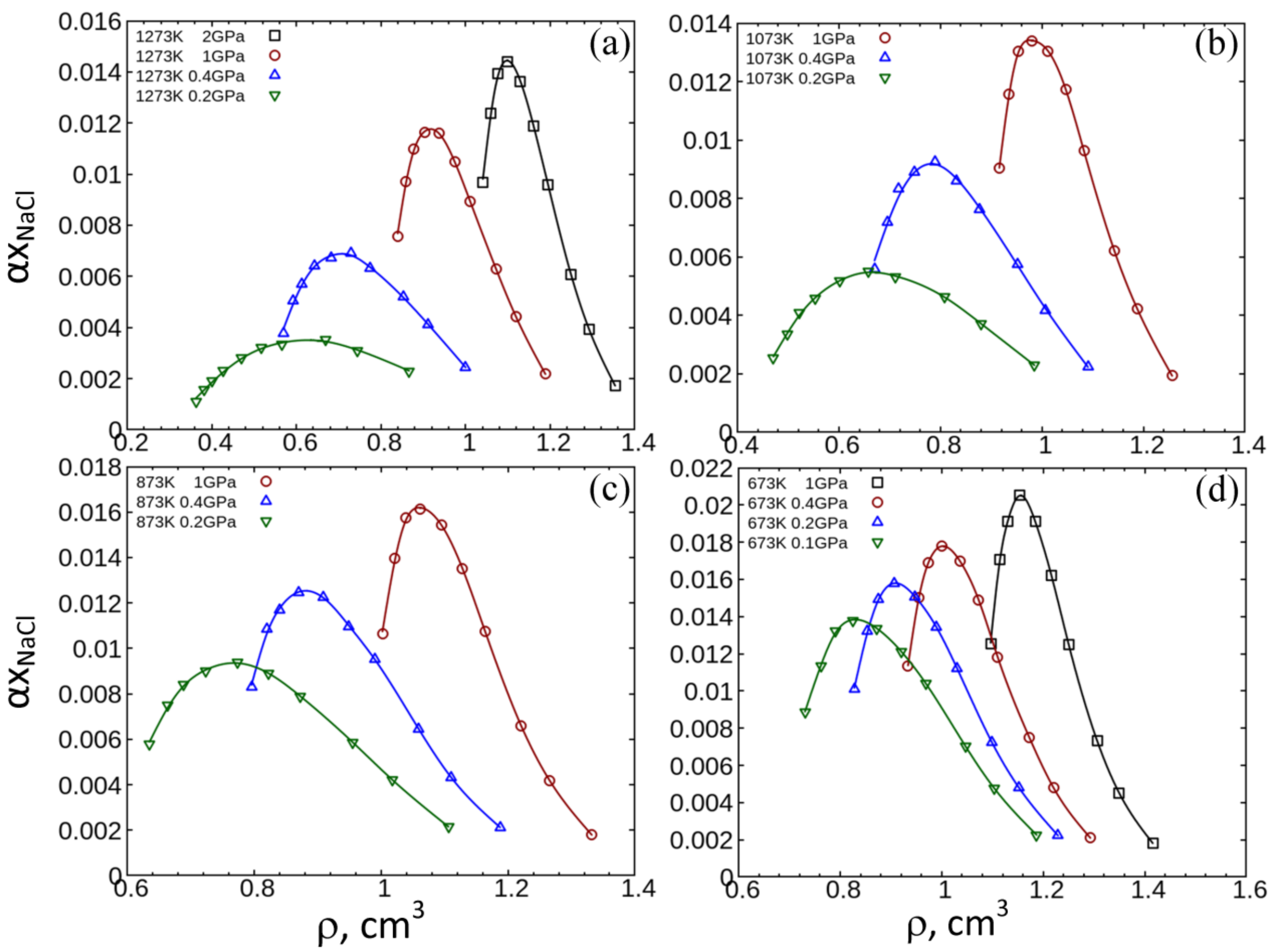

**FIG. 6.** Mole fraction of dissociated ion pairs as a function of the density of the fluid.

An increase in the spatial density of $Na^+$ and $Cl^-$ ions leads to the formation of ion clusters whose size exceeds that of the $Na^+Cl^-$ pair. As a numerical characteristic of this process, we calculated the average coordination number of ions associated with ions of the opposite charge $n_q$. Among $Na^+$ ions, we consider only those ions that are associated with at least one $Cl^-$ ion. For each $Na^+$ ion of this type, we count the number of $Cl^-$ ions at distances $r_{Na^+Cl^-} < q$. Averaging these numbers over $Na^+$ ions and the stored configurations of the system yields the value of $n_q$. This value coincides with $n_q$ obtained in a similar procedure for $Cl^-$ ions. In a dilute solution, where each bound ion is associated with a single ion of the opposite charge, $n_q = 1$. In the NaCl crystal, $n_q = 6$. In Fig. 7, the values of $n_q$ are presented as a function of the mole fraction of NaCl in the fluid. At the minimum salt concentration $x_{NaCl} = 0.018$, most curves, except those corresponding to low pressures and hence low water densities, do not exceed $n_q = 1.3$. Comparing the values of $n_q$ for the same pressure, for example $P = 1$ GPa, one can see that $n_q$ increases with temperature. This trend is consistent with data in Fig. 4(a), which shows a decrease in $K_d$ with increasing temperature.

An increase in $x_{NaCl}$ results in an increase in $n_q$ to $n_q \approx 3$ at $x_{NaCl} = 0.333$. The dependences $n_q(x_{NaCl})$ are close to linearity at the highest pressures shown for each temperature. Similar to the behavior shown in Fig. 2, a decrease in $n_q$ with increasing pressure in the region of relatively low $x_{NaCl}$ values is replaced, for high $x_{NaCl}$ values, by an increase in $n_q$ with increasing pressure. The behavior of the curve at $P = 0.2$ GPa and T = 1273.15 K is noteworthy. For low $x_{NaCl}$ values, this curve is clearly above the curves corresponding to higher pressures, indicating the preferential formation of multi-ion clusters at very low water densities. This point is considered in more detail in the Discussion section.

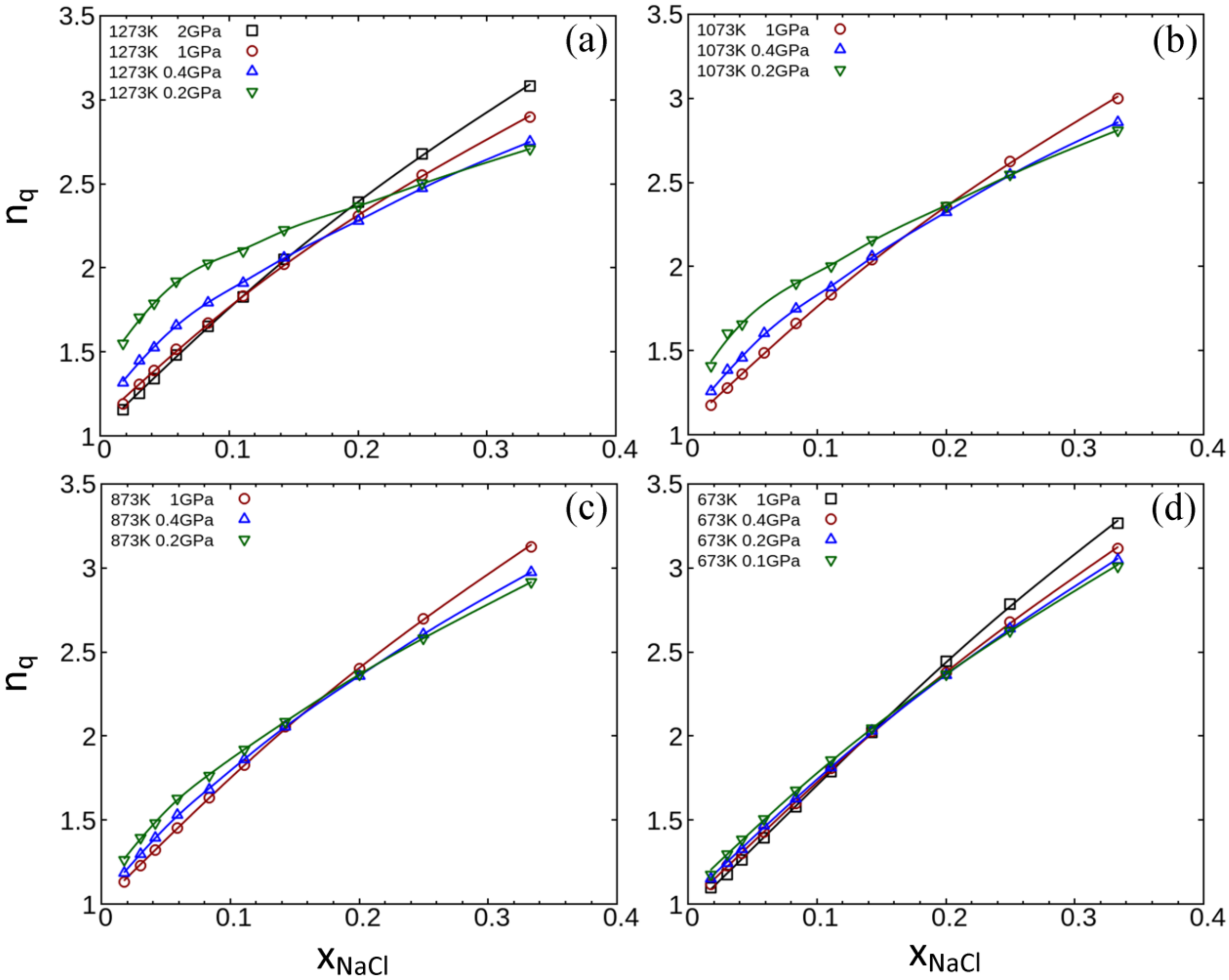

**FIG. 7.** Average coordination number of ions associated with oppositely charged ions as a function of the mole fraction of NaCl in the fluid.

### *3.2. Calculations at fixed volumes*

In addition to calculating $\log_{10}K_d$ and $\alpha$ at fixed pressure values, we performed calculations at fixed fluid densities and several temperature values for each density. The calculation results are presented in Table 1 for molality of 1m and in Table 2 for molality of 7m.

**Table 1**

Logarithm of the ideal dissociation constant, the degree of dissociation, and the pressure *P*, as a function of the fluid density *ρ* and the temperature at a fixed molality of 1m of the fluid $H_2O$–NaCl.

| 1m | $\rho = 0.5$ g/cm³ | | | $\rho = 0.6$ g/cm³ | | | $\rho = 0.7$ g/cm³ | | | $\rho = 0.8$ g/cm³ | | |
|---|---|---|---|---|---|---|---|---|---|---|---|---|
| *T*,K | $\log_{10}K_d$ | *α* | *P*,GPa | $\log_{10}K_d$ | *α* | *P*,GPa | $\log_{10}K_d$ | *α* | *P*,GPa | $\log_{10}K_d$ | *α* | *P*,GPa |
| 673.15 | -2.64 | 0.301 | 0.026 | -2.38 | 0.380 | 0.041 | -2.13 | 0.471 | 0.079 | -1.93 | 0.546 | 0.165 |
| 873.15 | -3.03 | 0.204 | 0.121 | -2.64 | 0.300 | 0.173 | -2.36 | 0.386 | 0.262 | -2.13 | 0.471 | 0.407 |
| 1073.15 | -3.22 | 0.167 | 0.220 | -2.80 | 0.258 | 0.309 | -2.49 | 0.345 | 0.443 | -2.26 | 0.421 | 0.645 |
| 1273.15 | -3.31 | 0.152 | 0.317 | -2.86 | 0.242 | 0.442 | -2.56 | 0.322 | 0.620 | -2.32 | 0.401 | 0.873 |
| 1473.15 | -3.30 | 0.154 | 0.411 | -2.89 | 0.236 | 0.570 | -2.58 | 0.316 | 0.790 | -2.34 | 0.394 | 1.094 |
| 1673.15 | -3.26 | 0.160 | 0.503 | -2.85 | 0.244 | 0.695 | -2.58 | 0.316 | 0.955 | -2.36 | 0.388 | 1.308 |
| **1m** | $\rho = 0.9$ g/cm³ | | | $\rho = 1.0$ g/cm³ | | | $\rho = 1.1$ g/cm³ | | | $\rho = 1.2$ g/cm³ | | |
| *T*,K | $\log_{10}K_d$ | *α* | *P*,GPa | $\log_{10}K_d$ | *α* | *P*,GPa | $\log_{10}K_d$ | *A* | *P*,GPa | $\log_{10}K_d$ | *α* | *P*,GPa |
| 673.15 | -1.76 | 0.615 | 0.323 | -1.62 | 0.668 | 0.595 | -1.53 | 0.700 | 1.021 | -1.42 | 0.743 | 1.654 |
| 873.15 | -1.95 | 0.539 | 0.638 | -1.78 | 0.604 | 0.991 | -1.69 | 0.640 | 1.513 | -1.61 | 0.671 | 2.257 |
| 1073.15 | -2.04 | 0.502 | 0.941 | -1.91 | 0.553 | 1.373 | -1.80 | 0.598 | 1.985 | -1.70 | 0.638 | 2.830 |
| 1273.15 | -2.12 | 0.474 | 1.234 | -2.00 | 0.520 | 1.739 | -1.87 | 0.568 | 2.434 | -1.79 | 0.602 | 3.381 |
| 1473.15 | -2.19 | 0.448 | 1.514 | -2.02 | 0.512 | 2.089 | -1.92 | 0.550 | 2.864 | -1.85 | 0.577 | 3.897 |
| 1673.15 | -2.19 | 0.447 | 1.783 | -2.05 | 0.499 | 2.425 | -1.95 | 0.540 | 3.277 | -1.89 | 0.563 | 4.400 |

The 1m case corresponds to the highest $\log_{10}K_d$ and $\alpha$ values obtained in this study. The density of the fluid listed in Table 1 varies from $\rho = 0.5$ g/cm$^3$ to $\rho = 1.2$ g/cm$^3$, and the temperature from $T = 673.15$ K to $T = 1673.15$ K. At temperatures between 673.15 and 1473.15 K and for all densities, the ideal dissociation constant and the degree of dissociation decrease with increasing temperature. Thus, the decrease of $\log_{10}K_d$ and $\alpha$ with temperature occurs at fixed fluid densities, while the decrease of $\log_{10}K_d$ shown in Fig. 4, has an additional source in the form of a decrease in density with temperature.

The temperature dependence of $\log_{10}K_d$ and $\alpha$ is most pronounced at low fluid densities and less pronounced at high $\rho$ values. The values of $\log_{10}K_d$ and $\alpha$ increase significantly with increasing $\rho$. This behavior can be associated with the increase in the dielectric constant of water and with stronger ion–water interactions in denser fluids. In the upper part of the temperature range shown and at $\rho = 0.5$–$0.6$ g/cm$^3$, a slight increase in $\alpha$ and $K_d$ with temperature can be observed.

At molality 7m, the $\log_{10}K_d$ values given in Table 2 are approximately 1–1.5 units lower than at molality 1m. The corresponding values for the degree of dissociation at 7m are about one order of magnitude lower than at 1m. As in the case of 1m, the values of $\alpha$ and $K_d$ increase with the density of the fluid at all the temperatures. At moderate temperatures, a decrease in these values with increasing temperature occurs for all the densities. At $\rho \leq 1.2$ g/cm$^3$, this decrease is followed by a slight increase in the higher temperature range.

**Table 2**

Logarithm of the ideal dissociation constant, the degree of dissociation, and the pressure *P*, as a function of the fluid density *ρ* and the temperature at a fixed molality of 7m of the fluid $H_2O$–NaCl.

| 7m | $\rho$ = 0.6 g/cm$^3$ | | | $\rho$ = 0.7 g/cm$^3$ | | | $\rho$ = 0.8 g/cm$^3$ | | | $\rho$ = 0.9 g/cm$^3$ | | |
|---|---|---|---|---|---|---|---|---|---|---|---|---|
| *T*,K | $\log_{10}K_d$ | $\alpha$ | *P*,GPa | $\log_{10}K_d$ | $\alpha$ | *P*,GPa | $\log_{10}K_d$ | $\alpha$ | *P*,GPa | $\log_{10}K_d$ | $\alpha$ | *P*,GPa |
| 673.15 | | | | | | | -3.02 | 0.088 | 0.011 | -2.84 | 0.108 | 0.077 |
| 873.15 | -3.70 | 0.041 | 0.075 | -3.38 | 0.059 | 0.108 | -3.14 | 0.078 | 0.176 | -2.99 | 0.092 | 0.305 |
| 1073.15 | -3.74 | 0.040 | 0.165 | -3.44 | 0.055 | 0.234 | -3.18 | 0.074 | 0.349 | -3.03 | 0.088 | 0.532 |
| 1273.15 | -3.69 | 0.042 | 0.258 | -3.40 | 0.058 | 0.362 | -3.20 | 0.073 | 0.520 | -3.04 | 0.086 | 0.758 |
| 1473.15 | -3.59 | 0.047 | 0.352 | -3.36 | 0.061 | 0.490 | -3.15 | 0.076 | 0.688 | -3.04 | 0.087 | 0.975 |
| 1673.15 | -3.51 | 0.052 | 0.445 | -3.29 | 0.066 | 0.615 | -3.12 | 0.079 | 0.854 | -3.01 | 0.090 | 1.188 |
| 7m | $\rho$ = 1.0 g/cm$^3$ | | | $\rho$ = 1.1 g/cm$^3$ | | | $\rho$ = 1.2 g/cm$^3$ | | | $\rho$ = 1.3 g/cm$^3$ | | |
| *T*,K | $\log_{10}K_d$ | $\alpha$ | *P*,GPa | $\log_{10}K_d$ | $\alpha$ | *P*,GPa | $\log_{10}K_d$ | $\alpha$ | *P*,GPa | $\log_{10}K_d$ | $\alpha$ | *P*,GPa |
| 673.15 | -2.71 | 0.124 | 0.223 | -2.62 | 0.137 | 0.488 | -2.57 | 0.145 | 0.914 | -2.54 | 0.149 | 1.555 |
| 873.15 | -2.84 | 0.108 | 0.524 | -2.76 | 0.118 | 0.872 | -2.70 | 0.125 | 1.395 | -2.68 | 0.129 | 2.142 |
| 1073.15 | -2.93 | 0.098 | 0.819 | -2.84 | 0.108 | 1.245 | -2.80 | 0.113 | 1.856 | -2.78 | 0.115 | 2.702 |
| 1273.15 | -2.95 | 0.095 | 1.106 | -2.88 | 0.103 | 1.604 | -2.83 | 0.109 | 2.297 | -2.80 | 0.112 | 3.237 |
| 1473.15 | -2.94 | 0.097 | 1.384 | -2.88 | 0.103 | 1.949 | -2.85 | 0.106 | 2.721 | -2.83 | 0.108 | 3.751 |
| 1673.15 | -2.92 | 0.098 | 1.653 | -2.89 | 0.102 | 2.283 | -2.86 | 0.106 | 3.129 | -2.85 | 0.107 | 4.241 |

## 4. Discussion

The values of the association and dissociation constants of NaCl in supercritical aqueous fluids have been published in earlier works (Quist and Marshall, 1968; Ho et al., 1994). The thermodynamic dissociation constants *K* and association constants were obtained from experimental data on the electrical conductivity of dilute sodium chloride solutions (0.001–0.1m) and reported to the standard state with molarity 1M. The *K* values were provided for conditions related to fixed densities of pure water $\rho_w$. We obtained our values of $\log_{10}K_d$ for a set of conditions presented by Quist and Marshall (1968) and Ho et al. (1994) and compare them with the logarithms of the thermodynamic dissociation constant obtained from data (Quist and Marshall, 1968; Ho et al., 1994). The comparison is provided in Table 3. To reproduce these conditions, we first calculated the pressures associated with the reported values of $\rho_w$ and *T* using the IAPWS-95 equation of state for pure water (Wagner and Pruß, 2002). The values of *P* are presented in Table 3. The $L_{box}$ cell size and the number of the water molecules for MD calculations were chosen to provide the correct pressure at the molarity 1 mol/dm$^3$. The number of the $H_2O$ molecules ranged from 303 to 701. The number of the $Na^+Cl^-$ pairs was set to 16. The values of $x_{NaCl}$ corresponding to the number of the particles are also given in Table 3.

**Table 3**

Comparison of our results for the logarithm of the NaCl ideal dissociation constant with thermodynamic dissociation constants data from Quist and Marshall (1968) and Ho et al. (1994).

| **1M** | $\rho_w$ = 0.3 g/cm$^3$ | | | | | $\rho_w$ = 0.4 g/cm$^3$ | | | | |
|---|---|---|---|---|---|---|---|---|---|---|
| ***T*,K** | $\log_{10}K_d$[a] | $\log_{10}K$[b] | $\log_{10}K$[c] | *P*,GPa | $x_{NaCl}$ | $\log_{10}K_d$[a] | $\log_{10}K$[b] | $\log_{10}K$[c] | *P*,GPa | $x_{NaCl}$ |
| 673.15 | -2.53 | -4.11 | -4.43 | 0.0288 | 0.032 | -2.49 | -3.19 | -3.20 | 0.0312 | 0.031 |
| 873.15 | -3.68 | -4.78 | -5.23 | 0.0810 | 0.045 | -3.23 | –3.67 | -3.77 | 0.1076 | 0.038 |
| 1073.15 | -4.10 | | -5.70 | 0.1306 | 0.050 | -3.52 | | -4.13 | 0.1832 | 0.040 |
| **1M** | $\rho_w$ = 0.5 g/cm$^3$ | | | | | $\rho_w$ = 0.6 g/cm$^3$ | | | | |
| ***T*,K** | $\log_{10}K_d$[a] | $\log_{10}K$[b] | $\log_{10}K$[c] | *P*,GPa | $x_{NaCl}$ | $\log_{10}K_d$[a] | $\log_{10}K$[b] | $\log_{10}K$[c] | *P*,GPa | $x_{NaCl}$ |
| 673.15 | -2.46 | -2.28 | -2.39 | 0.0372 | 0.030 | -2.29 | -1.76 | -1.67 | 0.0560 | 0.028 |
| 873.15 | -2.88 | -2.79 | -2.77 | 0.1461 | 0.033 | -2.58 | -2.14 | -2.00 | 0.2082 | 0.029 |
| 1073.15 | -3.05 | | -3.08 | 0.2547 | 0.034 | -2.71 | | -2.29 | 0.3573 | 0.029 |
| **1M** | $\rho_w$ = 0.7 g/cm$^3$ | | | | | $\rho_w$ = 0.8 g/cm$^3$ | | | | |
| ***T*,K** | $\log_{10}K_d$[a] | $\log_{10}K$[b] | $\log_{10}K$[c] | *P*,GPa | $x_{NaCl}$ | $\log_{10}K_d$[a] | $\log_{10}K$[b] | $\log_{10}K$[c] | *P*,GPa | $x_{NaCl}$ |
| 673.15 | -2.10 | -1.18 | -0.97 | 0.1050 | 0.025 | -1.91 | -0.73 | | 0.2106 | 0.022 |
| 873.15 | -2.27 | -1.51 | -1.67 | 0.3110 | 0.025 | -2.09 | -0.73 | | 0.4803 | 0.022 |
| 1073.15 | -2.44 | | | 0.5081 | 0.025 | -2.22 | | | 0.7332 | 0.022 |

[a] current work, [b] (Ho et al., 1994), [c] (Quist and Marshall, 1968)

Both our values of $\log_{10}K_d$ and $\log_{10}K$ from Quist and Marshall (1968) and Ho et al. (1994) increase with $\rho_w$ and decrease with temperature. There is some similarity in values of $\log_{10}K_d$ and $\log_{10}K$ at $\rho_w = 0.5$ g/cm$^3$ and $\rho_w = 0.6$ g/cm$^3$. However, the increase of $\log_{10}K$ with $\rho_w$ predicted in works (Quist and Marshall, 1968; Ho et al., 1994) occurs much faster than we obtain for $\log_{10}K_d$ in our MD calculations. The discrepancy may arise from several sources including the activity-coefficient corrections present in $K$ and absent in $K_d$, differences between conductometric and our structural definitions of ion association, standard-state conventions, force-field limitations, and extrapolation of the dilute-solution data.

**Table 4**

Our results on the degree of dissociation of NaCl compared with the data by Elbers et al. (2021).

| *T*, K | *ρ*, g/cm$^3$ | $N_{NaCl}$:$N_{H2O}$ | *α* (Elbers) | *α* |
|---|---|---|---|---|
| 673.15 | 0.720 | 5:100 | 0.26 | 0.257 |
| | | 20:400 | | 0.239 |
| 773.15 | 0.616 | 5:100 | 0.21 | 0.172 |
| | | 20:400 | | 0.159 |
| 873.15 | 0.616 | 5:100 | 0.22 | 0.154 |
| | | 20:400 | | 0.140 |

The association of the $Na^+Cl^-$ pairs in aqueous fluids, including the supercritical regime up to a temperature 600°C, was studied experimentally in the work (Elbers et al., 2021). Experiments carried out at molality $b$ = 2.63 mol/kg were supplemented by first principles MD calculations for a system consisting of 5 $Na^+Cl^-$ pairs and 100 $H_2O$ molecules (i.e. $b$ = 2.78 mol/kg). The results of the calculations were found to be in a good agreement with the experiment. We repeated the MD calculations for this system in the framework of our approach. The comparison of the results is presented in Table 4. Our calculations are presented here for both the system $N_{NaCl}$:$N_{H2O}$ = 5:100 and for a system four times larger, $N_{NaCl}$:$N_{H2O}$ = 20:400. We observe that the size of the first system is too small to neglect the influence of size. Our result for $T$ = 673.15 K with $N_{NaCl}$:$N_{H2O}$ = 5:100 is numerically consistent with the result by Elbers et al. (2021). These differences may be attributed to differences in the simulation methodology, force-field description, and statistical sampling, including the relatively limited number of configurations (100 samples) used in the cited study (Elbers et al., 2021).

MD calculations based on the SPC/E water model and ion parameters of (Smith and Dang, 1994) for concentrated aqueous solutions of NaCl (0.5–16m) at (25°C, 0.1MPa), (325°C, 0.1GPa), and (625°C, 1.5GPa) were performed by Sherman and Collings (2002). They obtained the density of the aqueous NaCl solutions under given pressure and temperature conditions and discussed the decrease in the association of ions into $NaCl^0$ molecules and larger NaCl clusters with increasing pressure. They also concluded that increasing the stoichiometric concentration of NaCl leads to a gradual change in the properties of the $H_2O$–NaCl system from an aqueous solution to a hydrated molten salt. The same conclusion was also reached in a subsequent study (Fowler, Sherman, 2020). Both of these studies do not provide detailed quantitative data on the degree of NaCl dissociation or association. Our results on the temperature dependence of $\alpha$ (Fig. 4) (decreasing with increasing temperature) are in agreement with the statement by Fowler and Sherman (2020) that the coordination number of Na–Cl increases with increasing temperature.

Our temperature dependence of $\alpha$ at $P = 1$ GPa and molality 3.5m can be compared with data by Fowler and Sherman (2020) for 3m. At T = 800°C, the degree of dissociation is $\alpha = 0.228$, while the work (Fowler and Sherman, 2020) indicates the coordination number of Na–Cl to be 0.75 at 3m. This corresponds to $\alpha = 0.25$. The authors of the cited study also found that, in the high pressure range, the Na–Cl coordination number is weakly dependent on the pressure. This corresponds to the dependences $\alpha(x_{NaCl})$ in Fig. 2, which are close to each other at high pressures.

In addition to dissociated ions and neutral molecules NaCl, the fluid $H_2O$–NaCl can contain multi-ion clusters formed by more than two ions (Oelkers and Helgeson, 1990; 1991; 1993; Sharygin et al., 2002; Elbers et al., 2021). These clusters can be neutral, positively or negatively charged. The results of our MD calculations allow us to study the formation of both neutral and charged multi-ion clusters in the fluid. However, the subject of this study is the dependence of the concentration of free $Na^+$ and $Cl^-$ ions on the temperature, pressure, and concentration of the NaCl in the fluid. A detailed analysis of the multi-ion clusters, in particular the dependence of their statistics on the $x_{NaCl}$, requires separate studies. Below, we will briefly consider only the possible influence of the negatively charged clusters on the ability of the fluid to dissolve and transport metals.

At a molality of 1m, the number of charged multi-ion clusters is typically about an order of magnitude smaller than the number of dissociated single ions. This ratio between charged clusters and individual ions increases with increasing $x_{NaCl}$ and reaches its maximum in the range from $x_{NaCl} = 0.1$ to $x_{NaCl} = 0.14$. Then this value decreases in the range $x_{NaCl} = 0.14 – 0.3$. Under all investigated PT conditions, except for the low density case $T = 1273.15$ K, $P = 0.2$ GPa, the number of multi-ion charged clusters is less than the number of individual dissociated ions. The maximum value of the ratio of charged clusters to individual dissociated ions is in the range of 0.65 – 0.9.

The presence of the charged multi-ion clusters cannot significantly affect our conclusion based on the shape of the curves in Fig. 5, showing the mole fractions of NaCl that are most favorable for the dissolution and transport of metals in the Earth's crust and mantle. Even if we hypothetically assume that the negatively charged multi-ion clusters bind metal ions as efficiently as individual $Cl^-$ ions, the maxima of curves similar to those shown in Fig. 5 for the sum of the mole fractions of $Cl^-$ ions and negatively charged multi-ion clusters shift to the right by about $\Delta x_{NaCl} = 0.02$. For the range of NaCl mole fractions most suitable for dissolving metal ions from rocks with subsequent transfer, the above assumption leads to values of $x_{NaCl} = 0.06 – 0.12$.

The only exception to the above is the $x_{Cl-}(x_{NaCl})$ at $T = 1273.15$ K and $P = 0.2$ GPa. The density of water under these conditions is 0.33 g/cm$^3$. This results in a decrease in the permittivity and other aspects of the interaction between $H_2O$ and NaCl. At molality of 1m, the value of $\alpha$ is only 0.05. Under these conditions, the number of charged multi-ion clusters is 0.75 of the number of individual dissociated $Na^+$ and $Cl^-$ ions. In the range of $x_{NaCl} = 0.042$ to $x_{NaCl} = 0.143$, the number of negatively charged clusters exceeds the number of free $Cl^-$ ions. The maximum value of the ratio of charged clusters to dissociated individual ions is 1.25 at $x$NaCl = 0.111. However, such low-density fluids are expected to have limited significance for metal transport because the absolute concentration of negatively charged species remains low.

Our choice between associated and dissociated states is based on the distance of the $Cl^-$ ion from the nearest $Na^+$ ion and is consistent with the role of $Cl^-$ ions in the mobility of metals in the Earth's crust and mantle. The $Cl^-$ ion, separated from the $Na^+$ ions, is capable of forming metal-chloride complexes. The opposite situation, when a $Na^+$ ion is in close proximity, prevents the formation of stable metal-chloride complexes and is considered an associate state. In this definition, the associated state does not depend on the duration of time that the ions are in close proximity to each other. In nature, not all dissociated $Cl^-$ ions can form stable complexes with metal ions. Therefore, our results indicate an upper limit on the possible concentration of metal-chloride complexes. In this context, our results complement previous geological and thermodynamic descriptions of concentrated NaCl-bearing fluids, in which NaCl has often been treated as predominantly dissociated or described through averaged dissociation behavior (Aranovich and Newton, 1996, 1997; Manning, 2018). The concentration dependence of α obtained here provides an additional molecular-level constraint that may help refine such descriptions, especially for highly concentrated supercritical fluids.

## 5. Conclusions

We investigated the dependence of the degree of dissociation and ideal dissociation constant of NaCl on the stoichiometric mole fraction of NaCl from $x_{NaCl} = 0.018$ to $x_{NaCl} = 0.333$ (from 1m to 27.8m) in supercritical aqueous fluids $H_2O$–NaCl at fixed pressures from 0.1 GPa to 2 GPa at temperatures from 673.15 K to 1273.15 K. It was found that the degree of dissociation decreases from typical values $\alpha \approx 0.3$–$0.7$ at the lower end of the concentration range to $\alpha \approx 0.004$–$0.007$ at $x_{NaCl} = 0.333$. Our calculations allowed us to obtain the mole fraction of the dissociated ions as a function of the stoichiometric mole fraction of NaCl in the fluid. In most cases, this mole fraction has a clearly defined maximum near $x_{NaCl} = 0.06$, which shifts to $x_{NaCl} = 0.10$ in low-density fluids. Due to the role of dissociated $Cl^-$ ions in the formation of metal-chloride complexes, supercritical $H_2O$–NaCl fluids in the salinity range close to this maximum may be relevant to conditions favorable for chloride-assisted metal transport. On the contrary, variations in the salt concentration, in particular its increase above $x_{NaCl} = 0.15 – 0.2$, can either sufficiently reduce the chloride-assisted transport capacity or lead to precipitation of metals and formation of ore occurrences.

Our calculations were carried out within the framework of the classical molecular dynamics using the flexible SPC/fw model of the water molecules and the Smith–Dang force field for the ion-ion and ion-water interactions. The numerical values obtained in our calculations can be the subject of more precise numerical or experimental studies. However, the existence of the effect of a significant decrease in the degree of dissociation in concentrated solutions does not depend on the method of calculations and has a geometric origin, i.e., in a concentrated solution there is not enough space to distribute a large number of ions at large distances from ions of the opposite charge.

Our definition of the dissociated state of $Na^+Cl^-$ pairs is based on the ability of the $Cl^-$ ions to act as ligands that form complexes with metals. The opposite, associated state implies the distance between $Na^+$ and $Cl^-$ ions is less than the position of the first minimum of the radial distribution function and is independent of the time spent in this immediate proximity. This associated state prevents the participation of the $Cl^-$ ions in dissolution and transport of metals by forming ligand complexes with them.

Possible further studies may include investigating the formation of multinuclear clusters in the studied range of NaCl concentrations and concentrations leading to the formation of macroscopic crystals, as well as investigating the properties of water in moderately and highly concentrated brines. Studies similar to those presented above can be carried out for other electrolytes, including aqueous solutions of non-1:1 electrolytes, such as $H_2O$–$CaCl_2$. $H_2O$–$CaCl_2$ is another common component of the geological fluids.

Possible wider implications could include a new analysis of geological data based on the found non-monotone dependence of the mole fraction of dissociated ion pairs on the mole fraction of the salt in the fluid. Knowledge of the dissociation degree of NaCl, which depends on temperature, pressure, and mole fraction of NaCl, is necessary for developing more realistic equations of state and thermodynamic models of multicomponent supercritical fluids. These models must take into account all these dependences of the dissociation degree of the participating electrolytes.

**Declaration of competing interest**

The authors declare that they have no known competing financial interests or personal relationships that could have appeared to influence the work reported in this paper.

**Acknowledgments**

The study was fulfilled under Research program of the IPGG RAS FMUW-2026-0005. The computational part of the work was performed using the equipment of the Shared Equipment Centre AIRES (Analytical Investigation of the Rising Earth Story) of the IPGG RAS.

**Author Contributions**

**Mikhail V. Ivanov:** Conceptualization (lead); Formal analysis (lead); Investigation (equal); Methodology (lead) ; Resources (equal); Software (equal) ; Supervision (lead); Validation (equal); Visualization (lead); Writing – original draft (lead); Writing – review & editing (equal). **Olga V. Olga V. Alexandrovich:** Conceptualization (supporting); Formal analysis (supporting); Investigation (equal); Methodology (supporting); Resources (equal); Software (equal); Validation (equal); Visualization (supporting) ; Writing – original draft (supporting); Writing – review & editing (equal).

**Data availability statement**

The derived data supporting the findings of this study are available within the article. The raw molecular dynamics trajectories and configuration data are available from the corresponding author upon reasonable request.